\documentclass[aps,prx,superscriptaddress,singlecolumn,longbibliography, notitlepage]{revtex4-1}

\usepackage{graphicx}
\usepackage{amsmath}
\usepackage{amssymb}
\usepackage{amsfonts}
\usepackage{amsthm}
\usepackage{hyperref}
\usepackage{color}
\usepackage{soul}
\usepackage{textcomp}
\usepackage{bm}
\usepackage{dsfont}
\usepackage{relsize}
\usepackage{braket}
\usepackage{enumitem}   
\usepackage{mathrsfs}
\usepackage{cleveref}
\usepackage{bigints}
\usepackage{framed}
\usepackage{comment}
\usepackage[makeroom]{cancel}

\theoremstyle{plain}

\numberwithin{obs}{section}

\newcommand{\comments}[1]{}
\newcommand{\ba}{\begin{align}}
\newcommand{\ea}{\end{align}}

\newcommand{\ee}{{\mathcal{E}}}

\newcommand{\bigo}[1]{\mathcal{O}\left (#1\right)}

\newcommand{\Tr}[1]{\text{Tr}\left[#1\right]}

\newcommand{\rom}[1]{\uppercase\expandafter{\romannumeral #1\relax}}
\newcommand{\norbra}[1]{\left( #1\right)}

\newcommand{\curbra}[1]{\left\{ #1\right\}}
\newcommand{\lind}{\mathscr{L}}
\newcommand{\J}{\mathbb{J}}

\def\be{\begin{eqnarray}}
\def\ee{\end{eqnarray}}

\begin{document}
	
	
	\title{Geometric optimisation of quantum thermodynamic processes}

	\author{Paolo Abiuso}
	
	\affiliation{ICFO – Institut de Ciències Fotòniques, The Barcelona Institute of Science and Technology, 08860 Castelldefels (Barcelona), Spain}
	
\author{Harry J.~D. Miller}
\affiliation{Department of Physics and Astronomy, The University of Manchester, Manchester M13 9PL, UK.}

\author{Mart\'i Perarnau-Llobet}
    \affiliation{D\' epartement de Physique Appliqu\' ee, Universit\' e de Gen\`eve, Gen\`eve, Switzerland}
	
		\author{Matteo Scandi}
	
	\affiliation{ICFO – Institut de Ciències Fotòniques, The Barcelona Institute of Science and Technology, 08860 Castelldefels (Barcelona), Spain}
	
	\begin{abstract}
	Differential geometry offers a powerful framework for optimising and characterising finite-time thermodynamic processes, both classical and quantum. 	Here, we start by a pedagogical introduction to the notion of thermodynamic length. We review and connect different frameworks where it emerges in the quantum regime: adiabatically driven closed systems, time-dependent Lindblad master equations, and discrete processes. A geometric lower bound on entropy production in finite-time is then presented, which represents a quantum generalisation of the original classical~bound. Following this, we review and develop some general principles for the optimisation of thermodynamic processes in the linear-response regime. These include constant speed of control variation according to the thermodynamic metric, absence of quantum coherence, and optimality of small cycles around the point of maximal ratio between heat capacity and relaxation time for Carnot~engines.
	\end{abstract}
	
		\maketitle


	\section{Introduction}

Quasistatic processes 
can be successfully characterised by a few simple and universal results: work is given by the equilibrium free energy difference between the endpoints of a transformation,  the~efficiency of a Carnot engine depends only on the temperatures of the thermal baths, and~ in general all  quantities of interest become state functions~\cite{landauStatisticalPhysics1980}.  These results are extremely strong, but~their applicability to real life situations is hindered by the necessity of performing all  protocols in infinite time in order to ensure that the system remains in thermal equilibrium along the process.   On~the other hand, finite-time thermodynamic processes can become incredibly complex and strongly depend on the particular protocol and system.  For~this reason,   universal results or simple characterisations are rare. A remarkable  exception are fluctuation theorems, which are universal results that apply to arbitrary out-of-equilibrium processes under very mild assumptions~\cite{Jarzynski2011}; however, they  provide a few constraints on the statistics, which are far from  sufficient for a full characterisation of the out-of-equilibrium process. 


	
	Noticeably, the~middle ground between the two situations above, i.e.,~the case in which the protocol is performed in long but finite time, can be characterised by few geometrical quantities. The~main ideas were introduced for classical systems in a series of seminal papers in the 80 s by Weinhold and Andresen, Berry and Salamon, among~ others~\cite{Weinhold, weinholdMetricGeometryEquilibrium1975b, Salamon3, Salamon1, Nulton1985, Schlogl1985, Andresen,RevModPhys.67.605,Hoffmann1989,diosi1996thermodynamic,andresen1996finite,sekimoto1997complementarity}. More recently, the~field saw a revival following a series of papers initiated by Crooks in 2007~\cite{Crooks,Zulkowski,Sivak2012a}, leading to several  applications in, e.g.,~ molecular motors~\cite{Sivak2016}, small-scale information processing~\cite{Zulkowski2015},  nonequilibrium steady
states~\cite{Zulkowski2013,Mandal2016a}, and~many-body systems~\cite{Rotskoff2015,Rotskoff2017}. The~same ideas have been generalised to the quantum regime for unitary dynamics using  linear response~\cite{Deffner2010,Deffner2013, Campisi2012geometric,Bonana2014,Ludovico2016adiabatic}, and~to open system dynamics for Lindbladian systems~\cite{DeWeese2015,scandi_thermodynamic_2019}.  Recent applications of thermodynamic geometry in quantum systems can be found in quantum heat engines~\cite{abiuso2020optimal,Brandner2020,bh2020geometric,hino2020geometrical}, equilibration processes~\cite{Mancino2018,Shiraishi2019},  phase transitions~\cite{Deffner2017}, quantum work and heat fluctuations~\cite{Miller2019,Scandi2020,miller2020quantum},   thermodynamic uncertainty relations~\cite{Guarnieri2019,miller2020thermodynamic}, and~ shortcuts to adiabaticity~\cite{acconcia2015shortcuts}; see also  Ref.~\cite{Deffner2020}  for a recent perspective on the~subject. 

	The goal of this paper is two-fold: First, we aim to provide a  pedagogic introduction to the notion of (quantum) \emph{thermodynamic length}. 
	This is done in Section~\ref{sec:thermolength_overview}, where we explicitly connect different frameworks where this concept can be derived: adiabatic linear response theory in closed quantum systems~\cite{Campisi2012geometric,Bonana2014,Ludovico2016adiabatic}, adiabatic Lindblad master equations~\cite{DeWeese2015,scandi_thermodynamic_2019}, and~discrete processes~\cite{Nulton1985}. Additionally,~
	in Section~\ref{sec:II}, we 
	use the concept of thermodynamic length to lower bound the dissipation in a finite-time process, generalising to quantum systems the so-called \emph{Horse--Carrot} theorem~\cite{Salamon1,Nulton1985}.  Notably, the~bound  is process-independent, being a function of the endpoints and the (smallest) relaxation timescale. Thus, it can be seen as a geometric refinement of the second law of thermodynamics. Second, in~Section~\ref{sec:optimisation}, we apply these ideas to the optimisation of thermodynamic processes, with~emphasis on heat engines in the low-dissipation regime~\cite{Salamon1,Schmiedl2007,esposito2010efficiency,Guo2013,VandenBroeck2013,Hernndez2015,Holubec2015,Holubec2016,ma2018universal,de2012optimal}. Building upon previous works, we show how general conclusions can be drawn with analytical tools for a class of thermal machines, and~a few principles of common application can be stated for optimal processes, with~some examples. 
	Finally, these results are illustrated in detail for the paradigmatic case of a finite-time Carnot engine with a driven two-level system as a working substance in  Section~\ref{sec:examplequbit}. 
 	
	
	\section{Overview of Thermodynamic Length in Quantum~Systems}
	\label{sec:thermolength_overview}

Let us consider a system whose Hamiltonian $H_t$ can be externally driven and which is weakly coupled to a thermal bath. Without~loss of generality, we will decompose the system Hamiltonian as $H_t = \sum_i \lambda_t^i X_i$, where $\curbra{\lambda_t^i}$ is a family of time dependent external parameters, and~$\curbra{X_i}$ are the corresponding observables. Moreover, in~the following we will assume summation over repeated indexes. In~this context the average work performed on the system is given by:
\begin{align}\label{eq:work1}
	w = \int_\gamma dt \; \Tr{\dot H_t \rho_t } = \int_\gamma dt \; \dot \lambda_t^i\:\Tr{X_i \rho_t },
	\end{align}
	where $\gamma$ is the path in the parameters space, and~$\rho_t$ is the evolved system density matrix at time $t \in (0, \tau)$. We know from equilibrium thermodynamics that if the process is infinitely slow the system is always at equilibrium. Consequently, the~work is given by the difference of free energy at the endpoints of the transformation. Indeed, in~this formalism we regain this result:
\begin{align}\label{eq:workEq}
	w_{\rm eq} = \int_\gamma dt \; \Tr{\dot H_t \pi_t } =   \int_\gamma dt \; \frac{d}{dt} \norbra{-\beta^{-1} \log \mathcal Z_t }  =  \Delta F  ,
	\end{align}
	where we used the notation $\mathcal{Z}_t = \Tr{e^{-\beta H_t}}$ for the partition function, we denote the thermal state by $\pi_t := e^{-\beta H_t}/\mathcal{Z}_t$, and~we used the definition of the free energy $F_t := -\beta^{-1} \log\mathcal{Z}_t$, as~well as $\Delta F = F_\tau - F_0$. Given this result, it is then natural to define the dissipated work as $w_{\rm diss} :=( w - w_{\rm eq}) =\norbra{w - \Delta F}$, in~order to isolate the role of the dissipation arising from finite time~effects. 
	
	A consequence of the second law is that $w_{\rm diss}\geq 0$ with equality only in the infinite time limit. Moreover, if~the dynamics is divisible (e.g., Markovian) the rate of dissipation is also  positive definite, and~zero only in the infinite time limit~\cite{Spohn1978}. This suggests that we can expand $\dot w_{\rm diss}$ in terms of $\{\dot \lambda_t^i\}$ around the quasistatic limit ($\dot \lambda_t^i\equiv 0$), and~obtain:
\begin{align}
	\dot w_{\rm diss} = \dot \lambda_t^i \cancel{\partial_i \; \dot w_{\rm diss}\big|_{\dot \lambda_t\equiv 0}} + \dot \lambda_t^i \norbra{\partial_i\partial_j \; {\dot w}_{\rm diss}\big|_{\dot \lambda_t\equiv 0}}  \dot \lambda_t^j +\bigo{|| \dot{\lambda} ||^3},
	\end{align}
	where the first derivative cancels since we are expanding around a minimum. For~the same reason, we~know that the Hessian $g_{i,j} =\beta \partial_i\partial_j \; \dot w_{\rm diss\big|_{\dot \lambda_t\equiv 0}}$ is positive definite. From~these considerations we see that the dissipated work can be written as:
\begin{align}\label{eq:generalMetric}
	w_{\rm diss} =\frac{1}{\beta }\int_\gamma dt \; \dot \lambda_t^i \; (g_{i,j})_t  \dot \lambda_t^j,
	\end{align}
	up to higher order corrections. Linear response theory tells us that the matrix $g_t$ depends smoothly on the thermal state $\pi_t$. Moreover, we can deduce that it is positive definite and symmetric, being~the Hessian of a function around its minimum. These are the defining properties of a metric. In~fact, we~can interpret Equation~\eqref{eq:generalMetric} as the energy functional or the action of the curve $\gamma$ with respect to the metric $g$. This name comes from the formal analogy between Equation~\eqref{eq:generalMetric} and the action of a system of free particles with mass tensor given by $g$. 
	
	This interpretation is particularly useful thanks to the following fact. If~one defines the length of $\gamma$ as:
\begin{align}\label{eq:generalDistance}
	l_\gamma = \int_\gamma dt \; \sqrt{\dot \lambda_t^i \; (g_{i,j})_t  \dot \lambda_t^j},
	\end{align}
	we have the Cauchy--Schwarz like expression
\begin{align}
	\label{eq:thermolength_ineq}
	\beta w_{\rm diss}\geq l_\gamma^2 / \tau ,
	\end{align}
	which takes the name of ``thermodynamic length inequality''~\cite{Salamon1}. Among~the curves connecting two~ endpoints, $\curbra{\lambda_0^i}$ and $\curbra{\lambda_\tau^i}$, we call $\gamma$ geodesic if it minimises the distance between the two points as measured by Equation~\eqref{eq:generalDistance}. A~geodesic is also characterised by the property that it keeps the product $\dot \lambda_t^i \; (g_{i,j})_t  \dot \lambda_t^j$ constant along its path, implying that the Cauchy--Schwarz inequality in Equation~\eqref{eq:thermolength_ineq} is saturated if $\gamma$ is a geodesic. 
	 Physically, this means that in order to design minimal dissipating protocols in the slow driving regime, 
	 it is sufficient to solve a system of differential~equations, i.e.,~the~geodesic~equations:
\begin{align}\label{eq:geodesicEquation}
\ddot \lambda^i_t +	\Gamma^i_{j,k}\big|_{\lambda_t} \,\dot \lambda^j_t\, \dot \lambda^k_t = 0,
\end{align}
where $\Gamma$ denotes the Christoffel symbols, which are given by:
\begin{align}\label{eq:chrisSymbols}
\hspace{-0.3cm}\Gamma^i_{j, k}|_{\lambda_t}  = \frac{1}{2} g^{i, l} \norbra{\partial_j g_{l, k} + \partial_k g_{j, l} - \partial_l g_{j, k}}|_{\lambda_t} .
\end{align}
Here, $g^{i, l}$ is the inverse of the metric, and~we use the shorthand notation $\partial_i g_{j, k}|_{\lambda_t}\equiv (\partial g_{j, k}/\partial \lambda_i)|_{\lambda = \lambda_t}$. 
	 Moreover, the~dissipative properties of a driven system can be directly inferred from the spectral properties of $g_t$ alone. In~particular, starting from very general considerations on the nature of the metric tensor, this will allow us to give lower bounds on the rate of dissipation (Section~\ref{sec:II}) and to conclude that the creation of coherence is always detrimental to the efficiency (Section~\ref{sec:coherence}). 
	
	Another strength of the formalism presented is that $g$ can be explicitly computed in many frameworks. For~example, comparing Equations~\eqref{eq:work1} and ~\eqref{eq:workEq} it can be seen that the metric tensor can be computed from the slow driving approximation of the expectation value of the observables $\curbra{X_i}$s. This~was~explicitly carried out in the context of linear response of an adiabatically driven unitary dynamics in~\cite{Ludovico2016adiabatic} (see also~\cite{Campisi2012geometric,Bonana2014}), leading to the expansion:
\begin{align}\label{eq:expLinResp}
	\Tr{X_i \rho_t }  = \Tr{X_i \pi_t } +  \chi_t^{\rm ad} [X_i, X_j] \dot \lambda_t^j + \bigo{||\dot\lambda ||^2},
	\end{align}
	where $\chi_t^{\rm ad}$ is the adiabatic response function given by:
\begin{align}\label{eq:chiAdiabatic}
	\chi_t^{\rm ad}[A, B] &= -i \int_{0}^\infty d\nu \;\norbra{ \nu \:\Tr{[A(\nu), B] \pi_t }}.
	\end{align}
	Here, we set $\hbar=1$, and~the Heisenberg picture $A(s)$ is defined with respect to the frozen Hamiltonian at time $t$, i.e.,~$A(s)=e^{i H_t s} A e^{-i H_t s} $. Notice that the upper bound of the integral can be extended to $\infty$ thanks to the exponential decay of the correlation function $\Tr{[A(\nu), B] \pi_t }$. Now, if~we plug the expansion just obtained in Equation~\eqref{eq:work1} and we recall that the definition of the dissipated work is $w_{\rm diss} :=( w - w_{\rm eq}) $, we have the expression:
\begin{align}
	w_{\rm diss} = \frac{1}{\beta }\int_\gamma dt\; \dot\lambda_t^i \,(\beta  \,\chi_t^{\rm ad} [X_i, X_j]) \dot \lambda_t^j,
	\end{align}
	up to higher order in $\{\dot \lambda_i\}$.   Comparing this equation with Equation~\eqref{eq:generalMetric}, we see that in the context of adiabatic linear response the metric tensor is given by $g^{\rm u}_{i,j} = \frac{\beta }{2}(\chi_t^{\rm ad} [X_i, X_j]+ \chi_t^{\rm ad} [X_j, X_i])$ (notice that even if $\chi_t^{\rm ad}$ is not in general symmetric in its arguments it can always be symmetrised without affecting the result, since  the velocities $\{\dot\lambda_t^i\}$ enter the integral in a symmetric way).  This formalism was recently used to geometrically characterise thermal machines close to Carnot efficiency~\cite{bh2020geometric}.
	
	Another relevant framework where a thermodynamic length can be derived is open quantum systems~\cite{scandi_thermodynamic_2019} (see also~\cite{DeWeese2015}). 
	In particular, consider  the
	Lindbladian dynamics:
\begin{align}
	\dot\rho_t = \lind_t[\rho_t],
	\end{align}
	with the property that each $\lind_t$ has the real part of all the eigenvalues negative and that there exist a unique instantaneous steady state $\pi_t$. These two conditions ensure that the dynamics asymptotically equilibrates irrespective of the initial conditions:
\begin{align}
	\lim_{\nu \rightarrow \infty} e^{\nu \lind_t} \rho = \pi_t.
	\end{align}
	In this case, it is possible to expand the state in the slow driving limit as $\rho_t \approx \pi_t + \delta \rho_t$~\cite{cavina2017slow}, where $\delta \rho_t$ can be expressed    up to higher order corrections as~\cite{scandi_thermodynamic_2019}:
\begin{align}
	\rho_t = \pi_t +\lind_t^+[\dot\pi_t] +\mathcal{O}(||\dot{\lambda}||^2),
	\end{align}
	where $\lind_t^+$ is the Drazin inverse of the Lindbladian given by:
\begin{align}
	\label{eq:def_drazin}
	\lind_t^+[A]  = \int_0^\infty d\nu\, e^{\nu\lind_t} \norbra{\pi_t \Tr{A} - A}.
	\end{align}
	As it will be shown explicitly in the following, the~eigenvalues of $\lind_t^+$ encode the information about the thermalisation timescales. Moreover, we introduce the shorthand notation to indicate the derivative of the state:
\begin{align}
	\dot\pi_t  =-\beta\, \dot\lambda_t^i\int_0^1 dx \;\pi_t^{1-x} \;\bar X_i \pi_t^{x} = -\beta\, \dot\lambda_t^i \J_t[\bar  X_i],
	\end{align}
	where we denote by $\bar X_i := X_i -\Tr{X_i\pi_t}$. Hence, if~we plug in this expansion into the expression of the work, we obtain that the dissipation takes the form:
\begin{align}\label{eq:dissOpensystem}
	w_{\rm diss} = -\frac{1}{\beta } \int_\gamma dt\; \dot\lambda_t^i \,(\beta^2 \, \Tr{\bar X_i \lind_t^+\J_t[\bar X_j]}) \dot \lambda_t^j.
	\end{align}
	Again, it should be noticed that the quadratic form $q_{i,j} = -\beta^2\, \Tr{\bar X_i \lind_t^+\J_t[\bar X_j]} $ is in general not symmetric, so that in the definition of the metric we need to explicitly symmetrise the expression: $g^d_{i,j} :=  \frac{1}{2}(q_{i,j}  + q_{j,i})$. The~matrix $g^d$ so defined can be then interpreted as the metric tensor for open quantum systems~\cite{scandi_thermodynamic_2019}. 
	
	It is interesting to notice that the metric $g^u$ obtained in the unitary setting can be cast in a form resembling the dissipative one $g^d$. In~fact, explicitly carrying out the integral in the definition of the adiabatic response function $\chi_t^{\rm ad}$, we see that the metric can be recast in the form:
\small{\begin{align}
	\chi_t^{\rm ad}[X_i, X_j] &= -i \int_{0}^\infty d\nu \;\norbra{ \nu \:\Tr{[X_i(\nu), X_j] \pi_t }}  =  -\frac{i}{ \mathcal{Z}_t}\int_{0}^\infty d\nu \;\norbra{ \nu \: e^{i (\varepsilon_m - \varepsilon_n)\nu}} (e^{-\beta \varepsilon_m}-e^{-\beta \varepsilon_n}) (X_i)_{m,n}(X_j)_{n,m}  \\
	&= -\frac{1}{ \mathcal{Z}_t} \frac{(e^{-\beta \varepsilon_m}-e^{-\beta \varepsilon_n})}{ (\varepsilon_m - \varepsilon_n)^2} (X_i)_{m,n}(X_j)_{n,m} = -i\beta\int_{0}^\infty d\nu \int_0^1 dx \; \Tr{\pi_t^{1-x} e^{i H_t \nu}X_i \:e^{-i H_t \nu}\pi_t^{x} X_j } \\
	&=-\beta\:\Tr{X_i \;\mathcal{U}_t^+[\J_t[ X_j]]},
	\end{align}}
	where we denoted by $\{\varepsilon_i\}$ the eigenvalues of $H_t$, and~we defined the operator:
\begin{align}
	\mathcal{U}_t^+[A] :=  -i \int_0^\infty d\nu\; {\rm Tr}_B [e^{-iH_t \nu} A e^{iH_t \nu}].
	\end{align}
	We see that the role of $\lind_t^+$ is taken in this case by the map $\mathcal{U}^+_t$, so that the dissipation in the unitary case is given in complete analogy to Equation~\eqref{eq:dissOpensystem}.

	One last example that one can consider is the case in which the Hamiltonian is changed in a sequence of quenches, followed by a perfect thermalisation of the system~\cite{Nulton1985}. The~total duration of the protocol is given by $\tau = N \tau_{\rm eq}$, where $N$ is the number of quenches in which the protocol is realised and $\tau_{\rm eq}$ is a fixed equilibration time. When the number of steps is large the state at each time $t=m\tau_{\rm eq}$ ($m=0,\dots,N-1$) is approximately given by: ${\rho_m \simeq \pi_m - \Delta_m \pi}$, where  $\Delta_m \pi$ is the difference between the thermal states at two subsequent steps $\Delta_m \pi:= \pi_{m+1} - \pi_{m}$. This term in the limit $N\gg 1$ is well approximated by $\tau_{\rm eq} \dot\pi_t$. We can interpret this contribution as an indication of how much the system lags behind the thermal state. Proceeding as before, the~dissipation can be rewritten up to first order in $1/N=\tau_{\rm eq}/\tau$ as:
\begin{align}\label{eq:BKM}
	w_{\rm diss} = \frac{1}{2\beta} \int_\gamma dt\; \dot\lambda_t^i \;(\tau_{\rm eq} \beta^2 \,\Tr{\bar X_i \J_t[\bar X_j]} )\dot \lambda_t^j.
	\end{align}
	The metric tensor $g^{q}_{i,j}$ can be directly identified with the trace inside the integral, since $\J_t$ is self-adjoint, making the whole expression symmetric in $(i,j)$. The~metric so obtained  can be rewritten as: \mbox{$g^{q}_{i,j} = \tau_{\rm eq} \,{\bf g}^{BKM}_{i,j}$}, where we implicitly defined ${\bf g}^{BKM}_{i,j} =\,\partial^2 \ln \mathcal{Z}/\partial \lambda_i \partial \lambda_j  $. 
	This last quantity is known as the Bogoliubov--Kubo--Mori (BKM) statistical distance, which encodes the geometry of the manifold of Gibbs states and has been thoroughly studied in the literature~\cite{PetzB, Petz2000, Petz, PetzG, Balian}. 
	Due to the formal similarity between \eqref{eq:BKM} and \eqref{eq:dissOpensystem}, it is insightful to study the relation between both metrics.   In~\cite{scandi_thermodynamic_2019}, it was shown that in the particular case in which the observables of interest $\curbra{Y_\alpha}$ are the left eigenoperators of the Lindbladian, meaning that they evolve according to the equation:
\begin{align}\label{eq:eigenops}
	\frac{d}{dt}\Tr{Y_\alpha \rho_t} = \tau_\alpha^{-1} \norbra{\Tr{Y_\alpha \pi_t}-\Tr{Y_\alpha \rho_t}},
	\end{align}
	 where $\curbra{\tau_\alpha}$ are the different timescales of the system, the~expression of the metric for the Lindbladian dynamics takes the simple form:
\begin{align}\label{eq:BKM2}
	g_{\alpha,\beta}^d = \frac{\tau_\alpha + \tau_\beta}{2} \;{\bf g}^{BKM}_{\alpha, \beta},
	\end{align}
 in analogy with the classical result~\cite{Sivak2012a}. 	Since, at~least for  Lindbladians satisfying detailed balance, $\{Y_\alpha\}$ is a complete basis of operators, it is possible to rewrite in this case any observable $X_i$ as $X_i = u_{i,\alpha}Y_\alpha$. That is, the~Lindbladian metric for a general family of observables $\{X_i\}$ is given by:
\begin{align}
		g_{i,j}^d = u_{i,\alpha} u_{j,\beta}\, \frac{\tau_\alpha + \tau_\beta}{2} \;{\bf g}^{BKM}_{\alpha, \beta}.
	\end{align}
	This shows that the role of $\lind_t^+$ is to encode the thermalisation timescales of the system, while the main geometrical properties are contained in ${\bf g}^{BKM}$. Finally, it should be noticed that in the case of a uniformly thermalising dynamics, i.e.,~$\tau_\alpha =\tau_{\rm eq} $  $\forall \alpha$,  the~thermodynamic metric is proportional to the BKM~one.
	
	\section{Bounding Dissipation with Thermodynamic~Length}	
	\label{sec:II}
	
	In a wider context, the~BKM metric plays a role within quantum information geometry~\cite{Hayashi2017b}, and~can be interpreted as a form of quantum Fisher information~\cite{Hayashi2002}. Moreover, it belongs to the family of contractive Riemann metrics over the manifold of normalised density operators $\varrho_t=\varrho_t(\{\lambda^i_t\})$. A~theorem by Petz gives a general characterisation of length between neighbouring quantum states~\cite{Petz1996a}:
\begin{align}\label{eq:metric}
	d\ell^2={\bf g}^f_{ij}d\lambda^i d\lambda^j \Longrightarrow {\bf g}^f_{ij}=\Tr{\frac{\partial \varrho_t}{\partial\lambda^i}c^f(R_{\varrho_t},L_{\varrho_t})\frac{\partial \varrho_t}{\partial\lambda^j}},
	\end{align}
	where $c^f(x,y)=(yf(x/y))^{-1}$ and $f(t)$ is a so-called Morozova--Cencov function which is operator monotone, normalised such that $f(1)=1$ and fulfils $f(t)=tf(1/t)$. Furthermore $L_\varrho,R_\varrho$ represent the left and right multiplication operators defined according to $L_\varrho [A]=\varrho A$ and $R_\varrho [A]=A\varrho $ respectively~\cite{Petz1996a}. For~each different metric we have a different notion of distance between density matrices over a path $\gamma$:
\begin{align}\label{eq:flength}
	\ell^f(\gamma):=\int_\gamma d\ell=\int_\gamma dt\;   \sqrt{{\bf g}^f_{ij}\dot{\lambda}^i \dot{\lambda}^j}.
	\end{align}
	For the particular choice $f(x)=(x-1)/\log x$ one obtains the BKM metric ${\bf g}^f_{ij}={\bf g}_{ij}^{BKM}$, namely
\begin{align}
	{\bf g}_{ij}^{BKM}=\int^1_0 dx \ \Tr{\bigg(\frac{\partial \log \varrho_t}{\partial\lambda^i}\bigg)\varrho_t^x\bigg(\frac{\partial \log \varrho_t}{\partial\lambda^j}\bigg)\varrho_t^{1-x}}.
	\end{align}
	Restricting to the manifold of thermal states $\varrho_t=\pi_t$ we indeed recover the thermodynamic metric in~\eqref{eq:BKM}. In~general, any length of the form~\eqref{eq:flength} is lower bounded by a geodesic path. Notably, analytical expressions for the shortest curves on the density operator manifold for each choice of metric are not known, aside from a couple of examples~\cite{Uhlmann1993,Gibilisco2013} excluding the BKM metric. However, for~the BKM statistical length a lower bound is known (Corollary 5.1 of~\cite{Jencova2004}) which depends only on the boundary conditions $\{\lambda^i_0\}\to \{\lambda^i_\tau\}$:
\begin{align}\label{eq:angle}
	\ell^{BKM}(\gamma)\geq \mathcal{L}(\varrho_{0},\varrho_{\tau}),
	\end{align}
	where
\begin{align}
	\label{eq:hellinger}
	\mathcal{L}(\rho,\sigma)=2 \arccos (\Tr{\sqrt{\rho}\sqrt{\sigma}}),
	\end{align}
	is the quantum Hellinger angle. We stress that while this bound can always be saturated when the initial and final states commute, transitions between non-commuting states cannot typically saturate~\eqref{eq:angle}. Note that in the classical commutative regime, all monotone metrics~\eqref{eq:metric} reduce to the classical Fisher--Rao metric, and~a unique geodesic length is singled out by the  Hellinger angle between the initial and final probability distribution~\cite{Gibilisco2013}. For~a pair of discrete classical probability distributions $p_n$ and $q_n$, the~Hellinger angle is given by
\begin{align}
	\label{eq:classical_hellinger}
	    \mathcal{L}(p,q):=2 \arccos \big(\sum_n \sqrt{p_n \ q_n}\big).
	\end{align}
	
	The geodesic bound~\eqref{eq:angle} has an immediate consequence for thermodynamics. For~step-equilibration processes, the~work dissipation~\eqref{eq:BKM} is subsequently lower bounded via the Cauchy--Schwartz inequality \eqref{eq:thermolength_ineq} combined with~\eqref{eq:angle}:
\begin{align}\label{eq:bound1}
	w_{\rm diss} \geq \frac{{k_B} T}{2N}\mathcal{L}^2(\pi_0,\pi_\tau).
	\end{align}
	One may interpret this as a geometric refinement to the second law of thermodynamics. Clearly,~the~bound depends only on the angle between the initial and final equilibrium state rather than the full path $\gamma$. For~open systems undergoing Markovian dynamics, the~corresponding dissipation~\eqref{eq:dissOpensystem} can be bounded in a similar fashion. Consider first the eigendecomposition of the Lindbladian~\eqref{eq:eigenops} with associated relaxation timescales $\{\tau_\alpha \}$, which can be achieved for open systems satisfying detailed~balance. Denoting $\tau_{\rm min}$ as the shortest timescale along the curve $\gamma$ and $\tau$ the total duration, work dissipation is bounded by
\begin{align}\label{eq:bound2}
	w_{\rm diss} \geq {k_B} T\bigg(\frac{\tau_{\rm min}}{\tau}\bigg)\mathcal{L}^2(\pi_0,\pi_\tau).
	\end{align}
    Note that, while \eqref{eq:bound1} can always be saturated by following a geodesic, in~general \eqref{eq:bound2} is not tight whenever more than one relaxation timescale is present. 
	The bounds~\eqref{eq:bound1} and~\eqref{eq:bound2} represent quantum generalisations of the so-called \textit{Horse--Carrot} theorem in finite-time thermodynamics~\cite{Salamon1,Nulton1985}.

	\subsection{Considerations on Coherence~Creation}\label{sec:coherence}
    Now we want to investigate the role of coherence in a a thermodynamic transformation whose dissipation can be described by Equation~\eqref{eq:dissOpensystem}, see also Refs.~\cite{Brandner2017,Scandi2020}. 
	We start by rewriting the expression for the dissipated work assuming full control on the system Hamiltonian
\begin{align}
	\dot{w}_{\rm diss}=-\beta \  \Tr{\dot{H}_t\lind_t^+\mathbb{J}_{\pi_t} \dot{H}_t}\equiv\langle\dot{H}_t,\dot{H}_t\rangle_t\ .
	\end{align}
	For notation simplicity we omit the explicit time dependence in this section. We split $\dot{H}$ in its diagonal and coherence parts, with~respect the Hamiltonian basis of $\pi\propto e^{-\beta H}$, $\ket{i}$
\begin{align}
	\dot{H}=\dot{H}^{(d)}+\dot{H}^{(c)} \qquad \dot{H}^{(d)}=\sum_i \ket{i}\bra{i}\dot{H}\ket{i}\bra{i}.
	\end{align}
	Given that for any operator $A$ we have $\Tr{A^{(d)} A^{(c)}}=0$, if~we are able to prove that $\mathbb{J}_\pi$ and $\lind^+$ do not mix the diagonal and coherent subspaces, then we would have
\begin{align}
	\label{eq:dissipation_d-c}
	\langle\dot{H},\dot{H}\rangle=\langle\dot{H}^{(d)},\dot{H}^{(d)}\rangle+\langle\dot{H}^{(c)},\dot{H}^{(c)}\rangle\ .
	\end{align}
	Now, this is always true for $\mathbb{J}_\pi$ as
\begin{align}
	\mathbb{J}_\pi[\ket{i}\bra{j}]=\int_0^1 dx \pi^x \ket{i}\bra{j} \pi^{1-x}\propto\ket{i}\bra{j}
	\end{align}
	meaning that if $\ket{i}\bra{j}$ is diagonal (i.e., $i=j$), it will stay diagonal, and~vice~versa (i.e., if~$i\neq j$).
	
	Is the same true for $\lind^+$? This question can be answered affirmatively, by~noting that $\lind^+$ can be written as an exponentiation of $\lind$ (cf. \eqref{eq:dissOpensystem}), and~that any $\lind$ satisfying detailed balance does not mix the diagonal and coherent subspaces~\cite{breuer2002theory}.  
	More explicitly, standard Markovian thermal Lindbladians (satisfying detailed balance~\cite{breuer2002theory,alickiDetailedBalanceCondition1976}) take  the form
	$
	\lind[\rho]=-i[H_{LS},\rho]+\sum_\alpha \gamma_\alpha A_\alpha\rho A_\alpha^\dagger-\frac{1}{2}\{A_\alpha^\dagger A_\alpha,\rho\} 
	$,
	the~$A_\alpha$ being jump operators $A_\alpha=\ket{i_\alpha}\bra{j_\alpha}$, and~$H_{LS}$ a general Lamb-Shift Hamiltonian~\mbox{$[H_{LS},H]=0$}. This commutation property guarantees that the Hamiltonian term does not mix populations with~coherences, while for the dissipative part we note
\begin{align}
	A_\alpha\ket{i}\bra{j} A_\alpha^\dagger-\frac{1}{2}\{A_\alpha^\dagger A_\alpha,\ket{i}\bra{j}\}=
	\ket{i_\alpha}\bra{i_\alpha}\delta_{j_\alpha i}\delta_{ j_\alpha j}-\frac{1}{2}\ket{i}\bra{j}(\delta_{j_\alpha i}+ \delta_{j_\alpha j})\ .
	\end{align}
	From the expression above, it is easy to see that if $i=j$ the result will be diagonal as well, while if $i\neq j$ the result will be only made of coherences. Equation~\eqref{eq:dissipation_d-c} is thus valid for standard Markovian master equations and
\begin{align}
	w_{\rm diss}= w_{\rm diss}^{(d)}+ w_{\rm diss}^{(c)}
	\end{align}
	where $w_{\rm diss}^{(d)}$ is the term due to the modification of the spectrum of $H$, while $w_{\rm diss}^{(c)}$ is due only to the rotation of the basis. Given that both $w_{\rm diss}^{(d)}$ and  $w_{\rm diss}^{(c)}$ are positive, this property immediately implies that $ w_{\rm diss}\geq w_{\rm diss}^{(d)}$, and~hence we conclude that the creation of coherence is always detrimental when operating a thermal machine in the low-dissipation regime, as~we explain more in detail in Section~\ref{subsec:path_optimisation}, and~in agreement with recent results~\cite{Brandner2017,menczel2020quantum,miller2020thermodynamic}. A~similar separation of losses generated by diagonal and coherent parts of the Hamiltonian variation is presented in~\cite{Brandner2020}.

	\section{Optimisation of Thermodynamic Processes in the Slow Driving~Regime}
	\label{sec:optimisation}
	
	In this section, we derive and review generic considerations on  the optimisation of  finite-time thermal machines in the low-dissipation regime~\cite{esposito2010efficiency,abiuso2020optimal,Salamon1,sekimoto1997complementarity, RefereeSuggestion}. That is, when the irreversible entropy production is proportional to the inverse time duration. This assumption can be taken as empiric if no information on the system--bath interaction is given, or~it can be justified and derived dynamically using the tools examined in Section~\ref{sec:thermolength_overview}. 
	Part of the results are in agreement with previous literature and we aim here to collect them in a unified exposition that shows the generality and simplicity hidden in earlier~works.

	More precisely, we consider a thermal machine made up of a working substance (or machine) and several thermal baths at different temperatures. The~level of control consists of  $n$ experimental parameters of the machine that can be driven (typically Hamiltonian parameters), together with the possibility to put the machine  in contact with one of the thermal baths. The~ $n$ control parameters are parametrised  as $\Vec{\lambda}(s)\equiv \Vec{\lambda}_{s\tau}$ with $s \in (0,1)$---note that this notation decouples the duration $\tau$ of each process from its shape $\Vec{\lambda}(s)$.  
	We assume in very general terms that the low-dissipation condition holds and it is described by an underlying thermodynamic metric, as~presented in Section~\ref{sec:thermolength_overview}. That is, for~an isothermal transformation at temperature $T=\beta^{-1}$, we rewrite Equation~\eqref{eq:generalMetric} as
\begin{align}
	\label{eq:optimisation_hypothesis}
	&\Delta Q = T\left(\Delta S - \frac{\sigma}{\tau}\right) \\ 
	&\sigma=\int_0^1 ds\;\vec{\lambda}'^T(s) g_{\vec{\lambda}} {\vec{\lambda}'}(s)
		\label{eq:optimisation_hypothesis2}
	\end{align}
	which follows from identifying $w_{\rm diss}=w- \Delta F= T\Delta S-\Delta Q= T\sigma/\tau$ and by recalling $\Vec{\lambda}(s)\equiv \Vec{\lambda}_{s\tau}$, which has derivative ${\vec{\lambda}'}\equiv\frac{\partial}{\partial s}\vec{\lambda}=\tau \dot{\vec{\lambda}}$.
	Notice that in most of what follows, the~exact form of $g_{\vec{\lambda}}$ does not significantly change the results.  In~this sense, most of the derivations are common to any system that has first-order losses described by some quadratic form, as~in linear response~theory.
	
		We consider a machine performing $M$ transformations close to equilibrium (in general with different baths), each described by some heat exchange and some dissipation in the low-dissipation regime, with~an output
\begin{align}
	\label{eq:general_LD_machine}
	\Delta W_{out}= \sum_i^M \Delta Q_i =\sum_{i=1}^M T_i\Delta S_i-\frac{T_i\sigma_i}{\tau_i}\ .
	\end{align}
	The output being a sum of heat exchanges is guaranteed when considering cycling machines, or~when the output of interest is the heat extraction from a subset of the sources. This framework thus includes a variety of tasks: cooling, work extraction, Landauer erasure, Carnot cycles, and~generalised Carnot engines with multiple baths or finite size baths (see examples below). In~any such a process,  three main features can be optimised, 
corresponding to different levels of control over the~machine:
	\begin{enumerate}
		\item {\bf The speed of the trajectory}: that is, the~duration $\tau$, which characterises the average speed of the process, plus any rescaling of the instantaneous velocity along the trajectory. This can be formalised as a change of coordinates $\vec{\lambda}(s)\rightarrow \vec{\lambda}(\mathfrak{s}(s))$ with $\mathfrak{s}$ smooth monotonous and $\mathfrak{s}(0)=0,\ \mathfrak{s}(1)=1$. 
		\item {\bf The path of the trajectory}: i.e.,~the (ordered) set of points swept by $\vec{\lambda}$, for~fixed $\vec{\lambda}(0)$ and $\vec{\lambda}(1)$. This identifies a curve $\gamma$ in $\mathbb{R}^n$.
		\item {\bf The extremal points} of $\gamma$, or~the ``location'' of the process in the control space.
	\end{enumerate}
	
	In the following, we elaborate on the above features and show how to optimise them, 
	which can be done independently or sequentially. In~particular, following the above order in Section \ref{subsec:time_optimisation} we~optimize the time duration of each transformation $\tau_i$ and show a principle of constant dissipation rate optimality; in Section \ref{subsec:path_optimisation} we discuss consequences of the considerations presented in Section~\ref{sec:II} when the experimental control is such to allow variations of the curve $\gamma$ defined by $\vec{\lambda}(s)$; and in \mbox{Section \ref{subsec:total_optimisation}} we discuss the cases in which a full optimisation can be carried out, so that all the degrees of freedom listed above can be~optimised. 

	\subsection{Tuning the Speed: Optimality of Constant Dissipation~Rate}
	\label{subsec:time_optimisation}
	Here, we suppose initially that the only control available on the machine~\eqref{eq:general_LD_machine} is the time tuning of each step $\tau_i$. We wish to maximise the power output $P=\Delta W_{out}/\sum_j \tau_j$ for a given loss, or~equivalently we fix the (maximum) amount of dissipated work,
\begin{align}
	\label{eq:dissip_constraint}
	\sum_i \frac{T_i\sigma_i}{\tau_i} \equiv w_{\rm diss}\,
	\end{align}
	and maximize $P$. 
	The power can be written as
\begin{align}
	\label{eq:genPow_LD}
	P=\dfrac{(\sum_i T_i\Delta S_i )- w_{\rm diss}}{\sum_j \tau_j}\ ,
	\end{align}
	hence, maximising it is equivalent to minimising $\sum_j \tau_j$ with the constraint~\eqref{eq:dissip_constraint}. This can be stated as
	\begin{center}
	\emph{Principle 0. Maximising the power at fixed dissipation is equivalent to minimising the dissipation at given~duration.}
	\end{center} 
	This remark is important as the main result of this subsection (the optimality of constant thermodynamic speed, or~dissipation rate) will thus be valid for all machines 
	performing  tasks that are limited by the above trade-off. Examples are: maximising the power, minimising the dissipation (or~entropy production) with fixed total time, or~hybrid figures of merit combinations, such as maximising the power with a fixed amount of total loss.  For~a discussion of what machines maximise their outputs when the irreversible entropy production is minimised see~\cite{salamon2001conditions}. 
	
	The maximisation of \eqref{eq:genPow_LD} can be done differentiating w.r.t $\tau_i$ and using Lagrange multipliers, or~directly with a Cauchy--Schwarz inequality
\begin{align}
	\label{ineq:CS_wdiss}
	w_{\rm diss}\sum_i\tau_i=\left(\sum_j \frac{T_j\sigma_j}{\tau_j}\right) \left(\sum_i \tau_i\right)\geq {\left(\sum_j \sqrt{T_j\sigma_j}\right)^2}
	\end{align}
	which is saturated when all $T_j\sigma_j/\tau_j^2$ are equal, that is
\begin{align}
	\label{eq:optimal_fixA}
	&\tau_j=\frac{\sqrt{T_j\sigma_j}(\sum_i \sqrt{T_i\sigma_i})}{w_{\rm diss}} \\
	&P_{w_{\rm diss}}=\frac{w_{\rm diss}(\sum_i T_i \Delta S_i) - w_{\rm diss}^2}{(\sum_j \sqrt{T_j\sigma_j})^2}\ .
	\end{align}
	Notice that the fact that $T_j\sigma_j/\tau_j^2$ is the same $\forall j$ means that the rate of dissipation is constant for each of the $N$ steps of the protocol.
	In particular, when the dissipation is described by an underlying thermodynamic metric~\eqref{eq:optimisation_hypothesis2}, this implies the optimality of constant thermodynamic velocity
	$   T\vec{\lambda}'^T g_{\vec{\lambda}} \vec{\lambda}'=const. $,
	which can be seen by dividing each transformation into infinitesimal steps, i.e.,~expressing
\begin{align}
	 T_i \Delta S_i-\frac{T_i\sigma_i}{\tau_i}
	=
	\int_{{\gamma}^{(i)}} TdS -\frac{T d\vec{\lambda}^T g_{\vec{\lambda}} d\vec{\lambda}}{d\tau}
	\end{align}
	and applying the above reasoning, which concludes that each of the infinitesimal $\frac{T d\vec{\lambda}^T g_{\vec{\lambda}} d\vec{\lambda}}{d\tau^2}$ must be equal. 
	The ``thermodynamic length inequality'' inequality \eqref{eq:thermolength_ineq} (\cite{Salamon1,andresen_constant_1994,diosi_thermodynamic_1996})
	is indeed saturated when its integrand is constant, and~coincides with the continuous version of~\eqref{ineq:CS_wdiss}.
	These considerations can be summed up saying that for the class of machines considered here
	\begin{center}
	\emph{Principle 1. In~optimal protocols, the~speed of the control variation is constant (as measured from the underlying thermodynamic metric), leading to a constant entropy production rate.}
	\end{center}
		The optimality of constant entropy production rate was noted already in the first seminal papers~\cite{salamon_minimum_1980} in the context of endoreversible engines, and~appeared in many works thereafter (for an historical perspective, see also~\cite{salamon_principles_2001,andresen_current_2011}).
	The above formulation manifests the universality of this principle whenever a trade-off between output rate and losses is present in the regime where losses are linear in the average speed of the~process.
	
	The power \eqref{eq:optimal_fixA} can be further maximised choosing $w_{\rm diss}=\frac{1}{2} \sum_i T_i\Delta S_i$ to obtain the durations leading to the maximum power, in~this case
\begin{align}
	\label{eq:maxPOW}
	{P}_{\rm max}=\frac{(\sum_i T_i\Delta S_i)^2}{4(\sum_j \sqrt{T_j \sigma_j})^2}\ .
	\end{align}
	At maximum power the losses thus correspond to half of the quasistatic output: this corresponds to the ``7th principle of control thermodynamics'' pointed out by Salamon~et~al. in~\cite{salamon_principles_2001}, whose general validity was unknown: we can state it holds (at least) for all machines described by~\eqref{eq:general_LD_machine}. 
	
	We give here an example of application of the time tuning optimisation just~described.
	
	\paragraph*{{Multi-Bath Carnot Engine.}} 
	\label{par:generalized_carnot_example}
	A generalised Carnot engine consists of a sequence of isotherms in contact with different thermal baths, alternated with adiabats as in the standard Carnot cycle. The~total work output can be expressed as the sum of the heat exchanges due to cycling conditions, as~in Equation~\eqref{eq:general_LD_machine}, with~
	$\sum_i \Delta S_i=0$. All the results described above apply and the maximum power obtainable by tuning the time durations of the isotherms is thus as in Equation~\eqref{eq:maxPOW}. Moreover, in~Appendix \ref{app:generalized_carnot} we further analyze this result assuming that all the baths have the same spectral density $\propto \omega^\alpha$, described by the ohmicity $\alpha$. Under~this hypothesis and the assumption that all the isotherms are small enough (see details in Appendix \ref{app:generalized_carnot}), we show how this can be translated in the maximum power being expressed by  
\begin{align}
	{P}^{\rm multi-Carnot}_{\rm max}=\frac{(\sum_i T_i dS_i)^2}{4\kappa_0 T_0\left(\sum_i (\frac{T_i}{T_0})^{\frac{1-\alpha}{2}}|dS_i|\right)^2} 
	\end{align}
	where $\kappa_0$ represents the local ratio between  $\sigma_0$ and $(\Delta S_0)^2$ at some reference temperature $T_0$, and~satisfies $\kappa_i/\kappa_j=(T_i/T_j)^{-\alpha}$. In~the Appendix \ref{app:generalized_carnot}, we show how in this case, the~power is upper bounded by the same power when it is obtained by the use of the highest and lowest temperature only, which leads to the maximum power of a standard Carnot Engine (cf. Section~\ref{subsec:total_optimisation} or~\cite{abiuso2020optimal})
\begin{align}
	\label{eq:generalized_carnot_final_main}
	{P}^{\rm multi-Carnot}_{\rm max}\leq {P}^{\rm Carnot}_{\rm max}=\frac{(\Delta S)^2}{\sigma_h}\frac{(T_h-T_c)^2}{4 T_h\left(1+(\frac{T_c}{T_h})^{\frac{1-\alpha}{2}}\right)^2}\ .
	\end{align}
	

	\subsection{Path Optimisation: Geodesics and~Coherences}
	\label{subsec:path_optimisation}
When the control over the working fluid allows not only to vary the speed of the transformation, but~includes possible modifications of the path $\gamma$ of the trajectory $\vec{\lambda}(s)$, the~machine can be substantially improved. 
The optimisation over $\gamma$ is independent from the time tuning considered in the previous section. It consists of finding the shortest path $\sigma=\int_{\gamma} {\vec{\lambda}}'^T g_{\vec{\lambda}} {\vec{\lambda}}'$
between two fixed points for each  isotherm \eqref{eq:optimisation_hypothesis2} considered in the cycle. Indeed, when the extremal points of a trajectory are fixed, the~quasistatic output is fixed and minimizing $\sigma$ always improves both power and the~efficiency. 
	
	More precisely,  with~the tools described in Section~\ref{sec:thermolength_overview}, each of the $\sigma_i$ in Equation~\eqref{eq:general_LD_machine} will be described as in \eqref{eq:generalDistance} by some metric $g^{(i)}$ and some trajectory $\vec{\lambda}_{(i)}$, in~the form
	$   \sigma_i=\int_{\gamma^{(i)}} {\vec{\lambda}}'^T_{(i)} g^{(i)}_{\vec{\lambda}} {\vec{\lambda}}'_{(i)}\ $.
	As mentioned earlier (see Section~\ref{sec:thermolength_overview} or Section~\ref{subsec:time_optimisation}), by~choosing the speed to be constant the above expression can be minimised to the thermodynamic length of the path $\gamma^{(i)}$
\begin{align}
	\label{eq:sigmai_length}
	\sigma_i=\left(\int_{\gamma^{(i)}}ds\; \sqrt{{\vec{\lambda}}'^T_{(i)} g^{(i)}_{\vec{\lambda}} {\vec{\lambda}}'_{(i)}}\right)^2\equiv l^2_{\gamma^{(i)}}\ .
	\end{align}
	This quantity depends only on the path $\gamma^{(i)}$ of the trajectory and not on its parametrisation $\vec{\lambda}(s)$, but~it can be further minimised by considering its minimum among all the possible paths linking the extremal points, which then defines the geodesics distance between the extremal points
\begin{align}
	\label{eq:sigmai_geod}
	d_{\vec{\lambda}(0),\vec{\lambda}(1)}=\min_{\substack{\gamma\ \text{with extremals }\\ \{\vec{\lambda}(0),\vec{\lambda}(1)\}}} l_\gamma
	\end{align}
	These considerations can be stated as follows:
	\begin{center}
	\emph{Principle 2. In~optimal protocols, the~driving minimises the entropy production, i.e.,~it follows a geodesic on the thermodynamic manifold.}
	\end{center}
	In the quantum case, as~showed in Section~\ref{sec:coherence}, the~irreversible entropy production can be split in two independent parts, one due to the variation of the spectrum $\dot{H}^{(d)}_t$ and one due to the rotation of the eigenvectors $\dot{H}^{(c)}_t$ of the Hamiltonian, i.e.,~$\dot{H}_t=\dot{H}^{(d)}_t+\dot{H}^{(c)}_t$ and
\begin{align}
	w_{\rm diss}= w_{\rm diss}^{(d)}+ w_{\rm diss}^{(c)}\ ,
	\end{align}
	where $w_{\rm diss}^{(X)}=-\beta \int dt\; \Tr{\dot{H}^{(X)}_t\lind_t^+\mathbb{J}_{\pi_t} \dot{H}^{(X)}_t}$, with~$X=d,c$. 
	Now, notice that the quasistatic (lossless) output of a thermal machine is given by the integral of the heat exchange, or~the work exchange, computed on the equilibrium state $\pi_t$, for~example
\begin{align}
	w_{\rm eq}= \int dt\; \Tr{\pi_t \dot{H}_t}=\int dt \; \Tr{\pi_t\dot{H}^{(d)}_t}\ ,
	\end{align}
	which shows how the work exchange only depends on the diagonal variation of $H$, that is the spectrum variation. This easily follows from the fact that for thermal states at temperature $T$ one has 
	$
	\Delta U= w + \Delta Q =w + T\Delta S\ ,
	$
	where all the quantities depend uniquely on the spectrum of the final and initial control $H_0$, $H_\tau$ (which define as well the spectrum of $\pi_0$, $\pi_\tau$). This means that given the most general control $H_t=U_tH^{(d)}_t U^\dagger_t$, where $H^{(d)}_t$ is diagonal in a time-independent basis, all the lossless heat and work exchanges are the same for the protocol in which only the spectrum is varied, $H^{(d)}_t$. At~the same time given $w^{(c)}_{\rm diss}\geq 0$, losses are clearly reduced using  $H^{(d)}_t$.
	From this we learn that, for~standard Markovian dissipators,
	\begin{center}
	\emph{Principle 3. Quantum coherences are not created in optimal protocols, i.e.,~non-commutativity $[H_t,H_{t'}]\neq 0$ is avoided.}
	\end{center}
	The effect of coherences inducing losses in the power was noted already in~\cite{Brandner2017} in the context of linear response theory of slowly driven engines with slowly driven temperature, and~more recently in~\cite{miller2020thermodynamic}. A~different approach to quantum dynamics, namely quantum jump trajectories, shows again the detrimental effects of coherence creation
~\cite{menczel2020quantum}. 
	Moreover, notice that if the degree of control on the thermal machine allows to eliminate any coherence creation, using commutative controls all the metrics defined in Equation~\eqref{eq:metric} collapse into the classical one and the geodesics distance between states is given by \eqref{eq:classical_hellinger},  and~the bound \eqref{eq:bound2} can be~saturated.

	We show here an example of application for a cooling process.
	\paragraph*{{Cooling/Work Extraction}} 
	\label{par:cooling_example}
	Suppose we are interested only in a subset of the heat currents that are part protocol, meaning that  relevant output is the heat extracted from one (or multiple) thermal sources, as~in a generalised refrigerator model. To~fix the ideas for a single bath to be cooled the cooling rate is
\begin{align}
	P^{\rm cooling}=\frac{ T_c \Delta S_c - \frac{T_c\sigma_c}{\tau_c}}{\tau_{ex} + \tau_c} \equiv \frac{T_c \Delta S_c - w_{\rm diss}}{\tau_{ex}+\tau_c}
	\end{align}
	where now $\tau_{ex}$ is additional time spent on parts of the cycle that do not contribute to the cooling output. The~optimisation for fixed loss $w_{\rm diss}$ applies as from \eqref{eq:optimal_fixA} leading to $ \tau_c=T_c\sigma_c/w_{\rm diss}\ ,$ and a power
\begin{align}
	{P}^{\rm cooling}_{w_{\rm diss}}=\frac{T_c \Delta S_c - w_{\rm diss}}{\tau_{ex}+T_c \sigma_c w_{\rm diss}^{-1}}\ ,
	\end{align}
	which clearly increases as $\sigma_c$ is minimised. 
	The overall maximum of the cooling rate becomes for a suitable choice of $w_{\rm diss}$
\begin{align}
	\label{eq:p_cooling_max}
	P^{\rm cooling}_{\rm max}
	=T_c\sigma_c\frac{\left(\sqrt{\Delta S_c\tau_{ex}/\sigma_c+1}-1\right)^2}{\tau_{ex}^2}=T_c\frac{\Delta S_c^2}{4 \sigma_c} - T_c\frac{\Delta S_c^3}{8 \sigma_c^2}\tau_{ex} + \mathcal{O}(\tau_{ex}^2)\ .
	\end{align}
	The above expressions are all decreasing in the value of $\sigma_c$, which is minimal when obtained on the geodesics of the transformation, as~from Equations~\eqref{eq:sigmai_length} and \eqref{eq:sigmai_geod}.
	For example, let us assume that the cooling consists of a single transformation from $\pi_x$ to $\pi_y$, with~no additional time $\tau_{ex}=0$, and~full control on the Hamiltonian defining $\pi_{x,y}=e^{-H_{x,y}/T_c}/\Tr{e^{-H_{x,y}/T_c}}$. Then,~ the~maximum cooling power is obtained for a coherence-free protocol $[H_x,H_y]=0$ that~leads to $\sigma_{\rm min}=2\tau_{\rm eq} \arccos (\Tr{\sqrt{\pi_x}\sqrt{\pi_y}})$ from \eqref{eq:hellinger}, whereas the maximum cooling rate is  obtained by substituting it into \eqref{eq:p_cooling_max}. If~the control does not allow for coherence-less transformations, or~the Lindbladian has several time-scales, upper bounds on the cooling rate can be obtained by the use of~ \eqref{eq:bound2}.

	\subsection{Choosing the Location: Total~Optimisation}
	\label{subsec:total_optimisation}
	After optimizing the time duration and trajectory of the transformations, the~resulting optimal output rates only depend on the end points of the transformations. The~final maximisation of such expressions is in general non-trivial. However, we note how the maximum power obtained in \eqref{eq:generalized_carnot_final_main} is proportional $\left(\Delta S\right)^2/\sigma$, which is maximal when  $\sigma$ takes the geodesics value described above \eqref{eq:sigmai_geod}. Thus,~this last quantity
\begin{align}
	\frac{\left(\Delta S\right)^2}{\sigma}=\frac{\left(S_{\vec{\lambda}(0)}-S_{\vec{\lambda}(1)}\right)^2}{d^2_{\vec{\lambda}(0),\vec{\lambda}(1)}}
	\end{align}
	can be maximised by changing the extremal of the transformation. The~same quantity appears as the leading term for the cooling rate in \eqref{eq:p_cooling_max}.
	We find this to be a strikingly general feature of all thermal machines whose dynamical information ultimately consists of just one simple isothermal transformation close to equilibrium. This is clearly the case for a single heat extraction from a bath as in \eqref{eq:p_cooling_max}, but~it happens also, e.g.,~for Carnot engines, which, due to the trivial dynamics at the quenches, have all relevant quantities which can be expressed solely in terms of the two isotherms. For~example, power and efficiency of a Carnot engine read:
\begin{align}
	&P^{\rm Carnot} = \frac{\Delta S( T_h-T_c)-\left(\frac{T_c \sigma_c}{\tau_c}+\frac{T_h \sigma_h}{\tau_h}\right)}{\tau_c+\tau_h},
	&\eta=\frac{Q_h+Q_c}{Q_h}
	=1-\frac{T_c (\Delta S+\frac{\sigma_c}{\tau_c})}{T_h (\Delta S-\frac{\sigma_h}{\tau_h})}\ ,
	\end{align}
	where $\Delta S$ is the variation of entropy during the hot isotherm, and~the irreversible entropy productions are proportional to each other on optimal protocols $\sigma_h/\sigma_c=(T_c/T_h)^{-\alpha}$, according to the spectral density of the baths~\cite{cavina2017slow,abiuso2020optimal} (cf. Appendix~\ref{app:generalized_carnot}). The~two isotherms are thus \emph{symmetric}, in~the sense that by construction they have an opposite entropy variation $\Delta S_h=-\Delta S_c$, and~the trajectories follow the same geodesics to link the endpoints~\cite{cavina2017slow,abiuso2020optimal}. After~time optimisation on $\tau_c,\tau_h$ in such a case it is clear from dimensional analysis that the resulting power can only be proportional to $(\Delta S)^2/\sigma_h$  (or equivalently $(\Delta S)^2/\sigma_c$ due to proportionality) multiplied by a function with the dimension of~temperature. 
	
	In more detail, it has been shown recently~\cite{abiuso2020optimal}  that is possible to express the maximum power at any given efficiency $\eta=(1-\delta) \eta_C = (1-\delta) (1-T_c/T_h)$ for a Carnot engine (see also~\cite{Holubec2016,ma2018universal}). We report here for simplicity only on the case where $\alpha=0$, thus $\sigma_c=\sigma_h=\sigma$, as~\begin{align}
	\label{Pcarnot}
	P^{\rm Carnot}_{\delta}=\frac{\left(\Delta S\right)^2}{4\sigma}\frac{(T_h-T_c)^2\delta(1-\delta)}{(1-\delta) T_c+\delta T_h}
	\end{align}
	The importance of the term $(\Delta S)^2/\sigma$ was noted already in~\cite{Hernndez2015} as a natural unit of entropy over time, defining the performance of thermal machines in the low-dissipation regime for any trade-off between power and efficiency. 
	The equivalent optimisation for a refrigerator has been conducted in~\cite{holubec2020maximum}, where one has a cooling power and COP coefficient (this time $\Delta S$ is defined to be positive on the cold isotherm)
\begin{align}
	&P^{\rm Refrigerator} = \frac{\Delta S T_c-\frac{T_c \sigma_c}{\tau_c}}{\tau_c+\tau_h},
	&\varepsilon=\frac{Q_c}{|Q_h|-Q_c}=
	\frac{T_c\left(\Delta S-\frac{\sigma_c}{ \tau_c}\right)}{T_h\left(\Delta S+\frac{\sigma_h}{ \tau_h}\right)-T_c\left(\Delta S-\frac{\sigma_c}{\tau_c}\right)}\ ,
	\end{align}
	which leads to a maximum cooling power at given COP (again we report it for flat spectral density $\sigma_c=\sigma_h$, see~\cite{holubec2020maximum} for generalisations) $\varepsilon=(1-\delta)\varepsilon_C=(1-\delta)T_c/(T_h-T_c)$
\begin{align}
	P^{\rm Refrigerator}_\delta=\frac{(\Delta S)^2}{4\sigma}\frac{ T_c(T_h-T_c)\delta}{T_h-\delta T_c}\ .
	\end{align}

	Crucially, the~maximisation of the $(\Delta S)^2/\sigma$ term can always be obtained by the use of a Cauchy--Schwarz inequality~\cite{abiuso2020optimal}, that is noticing that
\begin{align}
	\label{maxDS}
	\frac{(\int dS)^2}{\int ds \vec{\lambda}'^T g_{\vec{\lambda}} \vec{\lambda}'}
	=
	\frac{\left(\int ds\ \vec{\partial} S_{\vec{\lambda}}\cdot \vec{\lambda}' \right)^2}{\int ds \vec{\lambda}'^T g_{\vec{\lambda}} \vec{\lambda}'}
	\leq
	\int ds\ \vec{\partial} S_{\vec{\lambda}}^T g^{-1}_{\vec{\lambda}} \vec{\partial} S_{\vec{\lambda}}
	\leq
	\max_{\vec{\lambda}}\ \vec{\partial} S_{\vec{\lambda}}^T g^{-1}_{\vec{\lambda}} \vec{\partial} S_{\vec{\lambda}}\equiv \max_{\vec{\lambda}}\ C(\vec{\lambda})
	\end{align}
	The upper bound in \eqref{maxDS} can  be saturated by performing an infinitesimal cycles around the point where $C(\vec{\lambda})$ is maximised. 
	In the meaningful case in which the observables $X_i$ decay with a well defined timescale $\tau_{\rm eq}$,  the~dissipation is described by the Kubo-Mori metric (see Section~\ref{sec:II}), and~$C(\vec{\lambda})$ is exactly the heat capacity of the system divided by the equilibration time, 
	leading to~\cite{abiuso2020optimal}:
\begin{align}
	\label{bphc}
	\frac{(\Delta S)^2}{\sigma}  \leq \max_G \frac{ \mathcal{C}(G)}{\tau_{\rm eq}}.
	\end{align}
	Here, $G=\beta H$ is the adimensional Hamiltonian, and~the thermal state and the heat capacity can be expressed as $\pi=e^{-G}/\Tr{e^{-G}}$ and $\mathcal{C}(G)=\Tr{G^2\pi}-\Tr{G\pi}^2$.
	In other words, 
	\begin{center}
	\emph{Principle 4.	In~order to optimise the power-efficiency trade-off, perform the finite-time Carnot cycle around the point where the ratio between heat capacity and relaxation time of the working medium is maximised.}
	\end{center}
	This general principle is illustrated in the next section for a two-level Carnot~engine.

	\section{Case Study: Finite-Time Qubit Carnot~Engine} 
	
	\label{sec:examplequbit}
	
	In what follows, we analyse the exactly solvable case of a  heat engine where the engine consists of a driven two-level system:
\begin{align}
	H(t) = E(t) \sigma_z.
	\end{align}
	We consider a finite-time Carnot cycle  where the working substance is sequentially connected with two thermal baths at different temperatures (see  details of the cycle in~\cite{abiuso2020optimal}), and~focus on   the low-dissipation regime where the results of  Section~\ref{sec:optimisation} naturally apply.  We model the relaxation with any of the two baths by an exponential decay to equilibrium with timescale $\tau_{\rm eq}$, $\Tr{H \dot{\rho}}= \tau^{-1}_{\rm eq}\Tr{H(\pi-\rho)}$, which~corresponds to   the so-called reset master equation. In~this case, the~thermodynamic metric is given by the KMB~metric. 
	
 	

	

\begin{figure}
		\centering
		\includegraphics[width=0.7\linewidth]{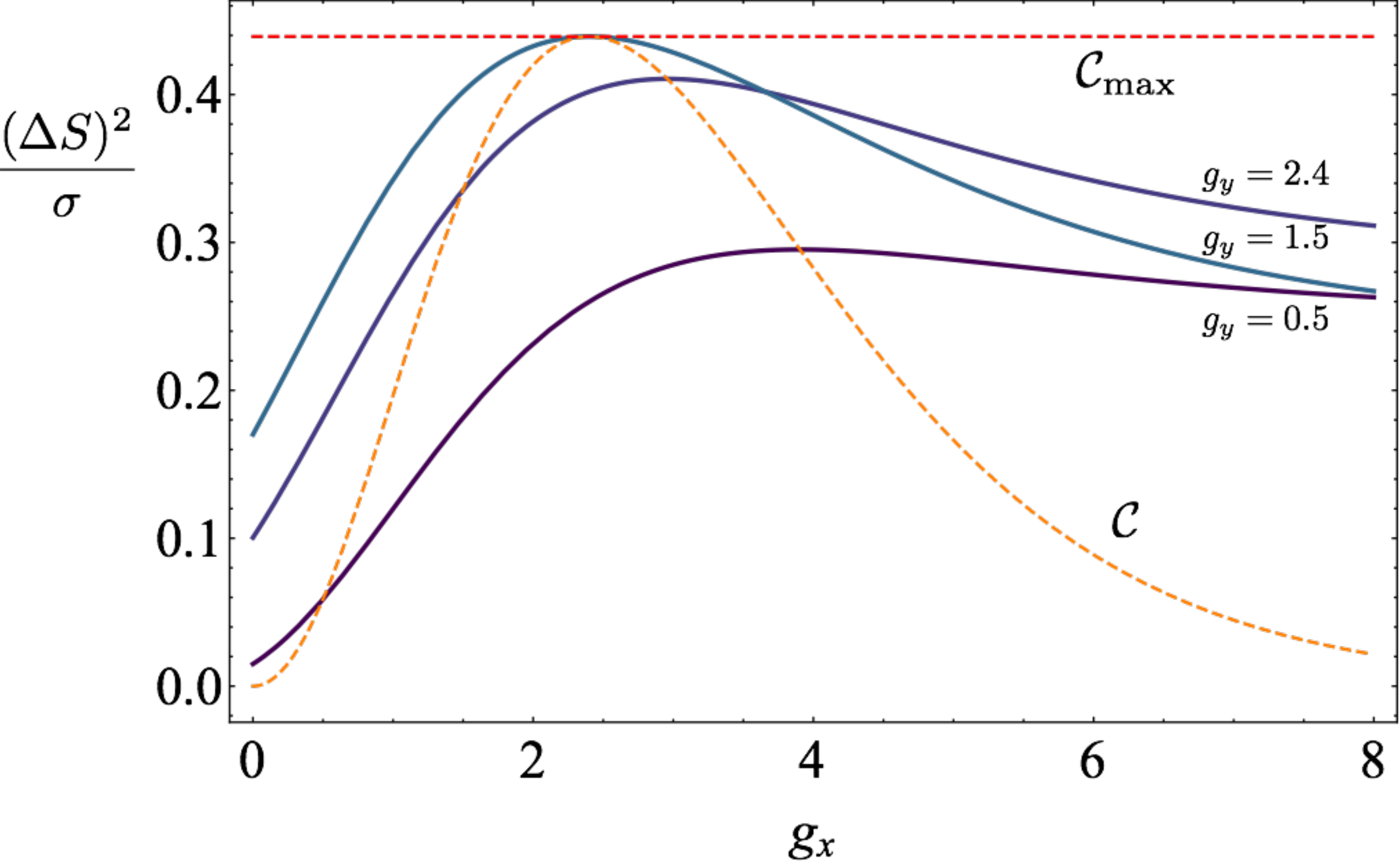} \caption{We plot the upper bound of $(\Delta S)^2/\sigma$, given in \eqref{upqubit}, as~a function of $g_x$ for different values of  $g_y=\{0.5,1.5,2.4\}$. The~point where $g_x=g_y\approx 2.4$ is the point where $(\Delta S)^2/\sigma$ is maximised (this can be easily checked numerically), which is also the point of maximum heat capacity $C$. The~heat capacity and its maximum are also plotted in dashed lines. We take $\tau_{\rm eq}=1$.  } 
		\label{fig1}
	\end{figure}

	
	Let us define $g \equiv \beta E$ (with $\beta$ being the inverse temperature of the bath the working substance is connected to), and~let $g_x$ and $g_y$ be the two endpoints of the isotherms, with~$g_x>g_y$. Let us also introduce the corresponding probabilities of the excited state:
\begin{align}
	&p_x =  \frac{e^{-g_x}}{1+e^{-g_x}},
	\nonumber\\
	&p_y = \frac{e^{-g_y}}{1+e^{-g_y}},
	\end{align}
	with $p_x<p_y$. 
	Then, we easily obtain: 
\begin{align}
	& \Delta S = - p_y \ln p_y - (1-p_y) \ln (1-p_y) +  p_x \ln p_x+  (1-p_x) \ln (1-p_x).
	\end{align}
	On the other hand, we can use \eqref{eq:bound2} to lower bound the entropy production in the isothermal processes as:
\begin{align}
	\sigma \geq  \tau_{\rm eq} \left(2 \arccos \left[\sqrt{p_x p_y}+ \sqrt{(1-p_x)(1-p_y)}   \right]\right)^2.
	\end{align}
	This bound can be saturated by following a geodesic, i.e.,~a protocol satisfying \eqref{eq:geodesicEquation}. 
	Putting~everything together, we can upper bound the relevant figure of merit $(\Delta S)^2 / \sigma$  for the power-efficiency optimisation as:
\begin{align}
	\label{upqubit}
	    \frac{(\Delta S)^2}{\sigma} \leq \frac{(- p_y \ln p_y - (1-p_y) \ln (1-p_y) +  p_x \ln p_x+  (1-p_x) \ln (1-p_x))^2}{\tau_{\rm eq} \left(2 \arccos \left[\sqrt{p_x p_y}+ \sqrt{(1-p_x)(1-p_y)}   \right]\right)^2}\;.
	\end{align}
	Importantly, this expression is protocol-independent and can be saturated. Indeed, the~maximal power of a finite-time Carnot engine (for a given efficiency $\eta=(1-\delta )\eta_C$)   given a two-level system can then be written as (see \eqref{Pcarnot}):
\small{\begin{align}
	{\rm max}_{\gamma} \hspace{1mm} P^{\rm Carnot}_{\delta}=\frac{1}{4} \frac{(- p_y \ln p_y - (1-p_y) \ln (1-p_y) +  p_x \ln p_x+  (1-p_x) \ln (1-p_x))^2}{\tau_{\rm eq} \left(2 \arccos \left[\sqrt{p_x p_y}+ \sqrt{(1-p_x)(1-p_y)}   \right]\right)^2}\frac{(T_h-T_c)^2\delta(1-\delta)}{(1-\delta) T_c+\delta T_h},
	\end{align}}
	where the maximisation is meant over all possible protocols in the slow driving regime.
	We show the upper bound \eqref{upqubit}  as a function of $g_x$ in Figure~\ref{fig1} for various values of $g_y$, including the optimal one, $g_y \approx 2.4$. It can be seen that the maximum of $(\Delta S)^2 / \sigma$ over $\{ g_x, g_y\}$ is bounded by the maximum of $\mathcal{C}/\tau_{\rm eq}$, where $\mathcal{C}$ is the heat capacity,
\begin{align}
	 	\mathcal{C} = g^2 p (1-p),
	 \end{align}
	 where $p$ is the excited state probability $p=e^{-g}/(1+e^{-g})$.
	This is in full agreement with \eqref{bphc} and~\cite{abiuso2019non}, and~is a particular illustration
	that the power of finite-time Carnot engines at any efficiency can be bounded  by substituting the maximum value of $\mathcal{C}/\tau_{\rm eq}$ to $(\Delta S)^2/\sigma$ inside expression~\eqref{Pcarnot}, as~discussed in detail in Ref.~\cite{abiuso2020optimal}. 
	
Summarising, here we have provided a tight upper bound on the relevant figure of merit $(\Delta S)^2 / \sigma$ for power (and efficiency) of a finite-time Carnot engine, for~the particular case of a  two-level driven~system. We note that such optimisation for a low-dissipation Carnot cycle or an Otto cycle has been performed in~\cite{abiuso2019non}, while exact total optimisation for a two-level system performing an arbitrary cycle was solved in Refs.~\cite{cavina2018optimal,erdman2019maximum},  with~both bosonic and fermionic baths. While our results apply  in the high efficiency or low-dissipation  regime, their strength lies in its simplicity: indeed, Equation \eqref{upqubit}  can be easily computed for  larger working substances, and~extensions to more complex relaxation processes with multiple timescales can also be relatively straightforwardly built (see Equation \eqref{maxDS} and Ref.~\cite{abiuso2020optimal}).  
This contrasts with exact results in finite-time thermodynamics~\cite{cavina2018optimal,Menczel2019b}, which rely on  non-trivial optimisation procedures that can become quickly unfeasible as the size of the working substance~increases. 



		\section{Conclusions and~Outlook}
		
	While originally developed for macroscopic systems, the~geometric approach to finite-time thermodynamics is now finding renewed applications within the emerging fields of stochastic and quantum thermodynamics. In~this paper, we have highlighted its utility for minimising dissipation in small scale systems operating close to~equilibrium. We have derived lower bounds on thermodynamic length that provide a geometric refinement to the second law of thermodynamics and allow one to benchmark the attainable efficiency of quantum thermal~machines. Alongside this, we summarised a set of key principles needed to optimise finite-time quantum low-dissipation engines in terms of efficiency and power, based on the computation of the thermodynamic metric tensor and length. Taken together, these principles provide a straightforward method for determining optimal thermodynamic processes. Indeed, we have seen that optimality is achieved by ensuring that the cycle follows a geodesic in the parameter space at constant velocity, while minimising the generation of quantum coherence and maximising the heat capacity relative to the relaxation time of the working~system.       
	
	Interesting future directions for thermodynamic geometry in the quantum regime 
	include the extension 
	beyond the slow driving regime~\cite{van2020geometrical}, the~minimisation and characterisation of work and heat~ fluctuations~\mbox{\cite{Miller2019,Scandi2020,denzler2020power,miller2020quantum}}, connections with strong coupling and speed-ups to isothermality~\cite{Pancotti2020},  application to cooling processes and relations with the third law of thermodynamics~\cite{Clivaz2019,Clivaz2019b,Guryanova2020},    many-body systems and criticality~\cite{Rotskoff2015,Rotskoff2017,Deffner2017}.







	
	

\vspace{6pt} 



\section*{Acknowledgements}{P.A.  is  supported  by ``la Caixa'' Foundation (ID 100010434, fellowship code LCF/BQ/DI19/11730023). M. P.-L. acknowledges funding from Swiss National Science Foundation (Ambizione  PZ00P2-186067). H. J. D. M. acknowledges support from the EPSRC through a Doctoral Prize. M.S. acknowledges funding from the European Union’s Horizon 2020 research and innovation programme under the Marie Sklodowska-Curie grant agreement No 713729. Both P. A. and M.S. also  acknowledge funding from Spanish MINECO (QIBEQI
FIS2016-80773-P, Severo Ochoa SEV-2015-0522), Fundacio Cellex, Generalitat de Catalunya (SGR 1381 and CERCA Programme).}

\appendix

\section{Optimality of Lowest-Highest Temperature Use in Multi-Bath Carnot~Engines}
\unskip

	\label{app:generalized_carnot}
	A generalised, finite-time Carnot engine between multiple thermal sources can be described as in Equation \eqref{eq:general_LD_machine} (where the adiabatic steps between the isotherms are assumed to happen on a much shorter timescale and thus neglected when compared to the $ \tau_i$s),
\begin{align}
	\Delta W_{out}= \sum_i^N \Delta Q_i =\sum_{i=1}^N T_i\Delta S_i-\frac{T_i\sigma_i}{\tau_i}\ ,
	\end{align}
	with $\Delta W\geq 0$ and where the index $i$ runs over multiple thermal baths, possibly with infinitesimal steps, including as a possibility the case in which the reservoirs have finite size~\cite{inprep_finite} and change temperature during the process (notice that in the case of finite size baths the total dissipation $\sum_i \frac{T_i\sigma_i}{\tau_i}$ is the natural measure of efficiency, as~the total work extractable from the machine sources is finite and obtainable in the quasistatic regime). 
	All the results of Section~\ref{subsec:time_optimisation} apply, and~the maximum power obtainable after tuning the $\tau_i$s can be written
\begin{align}
	{P}_{\max}=\frac{ (\sum_i T_i dS_i)^2}{4(\sum_j \sqrt{T_j\sigma_j})^2} .
	\end{align}
	To analyze further this result, we consider here the following property that holds for simple models where all the baths have the same spectral density
\begin{align}
	\label{eq:prop}
	\sigma_i=\kappa_{0} \left(\frac{T_i}{T_0}\right)^{-\alpha}dS_i^2
	\end{align}
	where $\alpha$ represents the spectral density exponent of the baths (their ohmicity), $T_0$ is a reference temperature that can be chosen at will, and~$\kappa_0$ a constant that depends on the local thermal state.
	This~property holds if the steps of the transformation are performed ``parallel'' to each other and are small enough for the state to be almost always the same. More precisely, baths with the same spectral density satisfy the property
\begin{align}
	    g_{H_1}^{(T_1)}=\left(\frac{T_1}{T_2}\right)^{-\alpha} g_{H_2}^{(T_2)} \quad \text{when} \quad \frac{H_1}{T_1}=\frac{H_2}{T_2}\ .
	\end{align}
	Here, $g$ is the metric that defines the dissipation in terms of the variation of $dG\equiv dH/T$ (cf.~Equation~ \eqref{eq:dissOpensystem}), and~the property $H_1/T_1=H_2/T_2$ means that the thermal state is the same $\pi_1=\pi_2$.  The~absolute value of the variation of entropy is instead the same if $dG_1=\pm dG_2$, as~in such a case
\begin{equation}
	    |dS_1|=|\Tr{d\pi_1 G_1}|=|\Tr{d\pi_2 G_2}|= |dS_2|\ .
	\end{equation}
	Combining the above two equations, we obtain \eqref{eq:prop}.
	For more details see~\cite{cavina2017slow} or the supplementary material of~\cite{abiuso2020optimal}.  
	For such a case we obtain substituting \eqref{eq:prop}
\begin{align}
	\label{eq:maxpow_generalizedCarnot}
	\bar{P}=\frac{(\sum_i T_i dS_i)^2}{4\kappa_0 T_0\left(\sum_i (\frac{T_i}{T_0})^{\frac{1-\alpha}{2}}|dS_i|\right)^2} \ .
	\end{align}
	Moreover, for~a cycle we have $\sum_i dS_i=0$ and we can divide the $N$ steps into those having $dS_{k^+} > 0$ (which we will indicate with the index $k^+$ and those having $dS_{k^-} < 0$ (with index $k^-$). We have thus $\sum_{k^+} dS_{k^+} = -\sum_{k^-} dS_{k^-}\equiv \mathcal{S}$. The~power \eqref{eq:maxpow_generalizedCarnot} can then be expressed in terms of the ``weights'' associated to each step for the positive and negative entropy variations. That is, we define
\begin{align}
	\label{eq:pi_dS}
	p_{k^+}=\frac{dS_{k^+}}{\mathcal{S}} \qquad p_{k^-}=-\frac{dS_{k^-}}{\mathcal{S}}
	\end{align}
	The vectors $p_{k^+}$ and $p_{k^-}$ are normalised probability vectors and the power \eqref{eq:maxpow_generalizedCarnot} can be written as
\begin{align}
	\label{eq:barP_prelemma}
	4\kappa_0 T_0\bar{P}=\frac{(\sum_{k^+} T_{k^+}p_{k^+} - \sum_{k^-} T_{k^-}p_{k^-})^2}{\left(\sum_{k^+} (\frac{T_{k^+}}{T_0})^{\frac{1-\alpha}{2}}p_{k^+}+ \sum_{k^-} (\frac{T_{k^-}}{T_0})^{\frac{1-\alpha}{2}}p_{k^-}\right)^2}
	=\left(\frac{\vec{T}_+\cdot \vec{p}_+ -\vec{T}_-\cdot \vec{p}_-}{\vec{T}'_+\cdot \vec{p}_+ +\vec{T}'_-\cdot \vec{p}_-}\right)^2
	\end{align}
	where we defined 4 positive vectors $\vec{T}_+,\vec{T}_-,\vec{T}'_+,\vec{T}'_- > 0$. Being allowed to modify separately we positive and negative weight (essentially by tuning the size of the entropy variations \eqref{eq:pi_dS}) it is possible to maximize the above quantity by noting that for any probability vector $\vec{p}$, positive vectors $\vec{B}>0$, vector $\vec{C}$, positive constant $b>0$, and~constant $c$, it holds
\begin{align}
	\label{eq:lemma_ineq}
	\frac{c+\vec{C}\cdot \vec{p}}{b+\vec{B}\cdot \vec{p}}\leq \max_i \frac{c+C_i}{b+B_i}
	\end{align}
	which is saturated by choosing $p_i=\delta_{i\bar{i}}$, where $\bar{i}$ is the index saturating the maximum of \eqref{eq:lemma_ineq}. Applying~twice the above inequality to $\sqrt{4\kappa_0T_0\bar{P}}$ of Equation~\eqref{eq:barP_prelemma} we obtain
\begin{align}
	\label{eq:barP_ineq}
	\sqrt{4\kappa_0T_0\bar{P}} \leq 
	\max_{ij} \frac{{T_+}_i-{T_-}_j}{{T'_+}_i+{T'_-}_j}\ .
	\end{align}
	Given that ${T'_\pm}_i={T_\pm}_i ^{\frac{1-\alpha}{2}}$, we study the function
\begin{align}
	f(x,y)= \frac{x-y}{x^\beta+y^\beta} \qquad x\geq y\geq 0
	\end{align}
	and find that it is always decreasing in $y$. Also, it increases always in $x$ provided that $\beta\leq 1$. We thus conclude that for $\alpha\geq -1$ the maximisation on the right-hand side of \eqref{eq:barP_ineq} is obtained by using the highest and lowest temperature available, that we will call $T_h$ and $T_c$ respectively.
	We thus find that
\begin{align}
	\label{eq:maxpow_generalizedCarnot}
	\bar{P}
	\leq 
	\frac{(T_h-T_c)^2}{4\kappa_0 T_0\left((\frac{T_h}{T_0})^{\frac{1-\alpha}{2}}+(\frac{T_c}{T_0})^{\frac{1-\alpha}{2}}\right)^2}
	\end{align}
	which is saturated when $dS_c=-dS_h$ and all the rest are null. This shows that under the assumption of equal spectral density the power is bounded by the power obtainable by using only the extremal~baths.


	
\end{document}